\documentclass{PoS}

\usepackage{cleveref}
\usepackage{wrapfig}
\usepackage{amsmath}
\usepackage{graphicx}
\usepackage[caption=false]{subfig}
\usepackage{multirow}
\usepackage{url}
\usepackage{wrapfig}
\usepackage{subfig}
\usepackage{float}

\title{The Scintillator Upgrade of IceTop: Performance of the prototype array}
\ShortTitle{Scintillator Performance}

\author{
The IceCube Collaboration\footnote{For collaboration list, see PoS(ICRC2019) 1177.}\\
{\itshape \href{http://icecube.wisc.edu/collaboration/authors/icrc19_icecube}{http://icecube.wisc.edu/collaboration/authors/icrc19\_icecube}}\\
E-mail: \email{mkauer@icecube.wisc.edu}
}

\abstract{

The IceCube Collaboration foresees to upgrade IceTop, the present surface array, with scintillator detectors augmented by radio antennas. As one of several goals the scintillator detectors will be used to measure and mitigate the effects of snow accumulation on the IceTop tanks: the increasing energy threshold and efficiency loss are nowadays the sources of the largest systematic uncertainties in shower reconstruction and mass composition analysis. In addition, the upgrade will provide useful experience for the development of next generation neutrino detectors proposed for the South Pole. In the Austral summer season, 2017-2018 two full ``stations'' were installed near the center of the IceTop array. Each station features custom-designed electronics and consists of seven detectors, each having an active area of 1.5m$^{2}$ plastic scintillator and wavelength shifting fibers read out by a Silicon Photomultiplier. In this contribution we review the detector design and performance, and show results from more than one year of operation of the prototype stations. During that year several thousand air shower events have been measured in coincidence with IceTop.\\

\vspace{4mm}
{\bfseries Corresponding authors:}
\speaker{Matt Kauer}$^{1}$, Thomas Huber$^{2,3}$, Delia~Tosi$^{1}$, Chris Wendt$^{1}$\\
{$^{1}$ \itshape Dept. of Physics and Wisconsin IceCube Particle Astrophysics Center, University of Wisconsin, Madison, WI 53706, USA}\\
{$^{2}$ \itshape Institut f\"{u}r Kernphysik, Karlsruhe Institute of Technology, D-76021 Karlsruhe, Germany}\\
{$^{3}$ \itshape DESY, D-15738 Zeuthen, Germany}
}

\FullConference{36th International Cosmic Ray Conference -ICRC2019-\\
		July 24th - August 1st, 2019\\
		Madison, WI, U.S.A.}

\begin{document}

\section{Introduction} 
\label{sec:Introduction}

IceCube is a cubic-kilometer neutrino detector installed in the ice at the geographic South Pole \cite{Aartsen:2016nxy} between depths of 1450\,m and 2450\,m, which was completed in 2010. Reconstruction of the direction, energy, and flavor of the neutrinos relies on the optical detection of Cherenkov radiation emitted by charged particles produced in the interactions of neutrinos in the surrounding ice. Additionally, an array of surface detectors, IceTop, has also been deployed for cosmic ray studies in the PeV energy range \cite{IceCube:2012nn} and to provide a partial veto of the down-going background of penetrating muons. 

\begin{figure}[ht]
    \centering
    \includegraphics[width=0.95\textwidth]{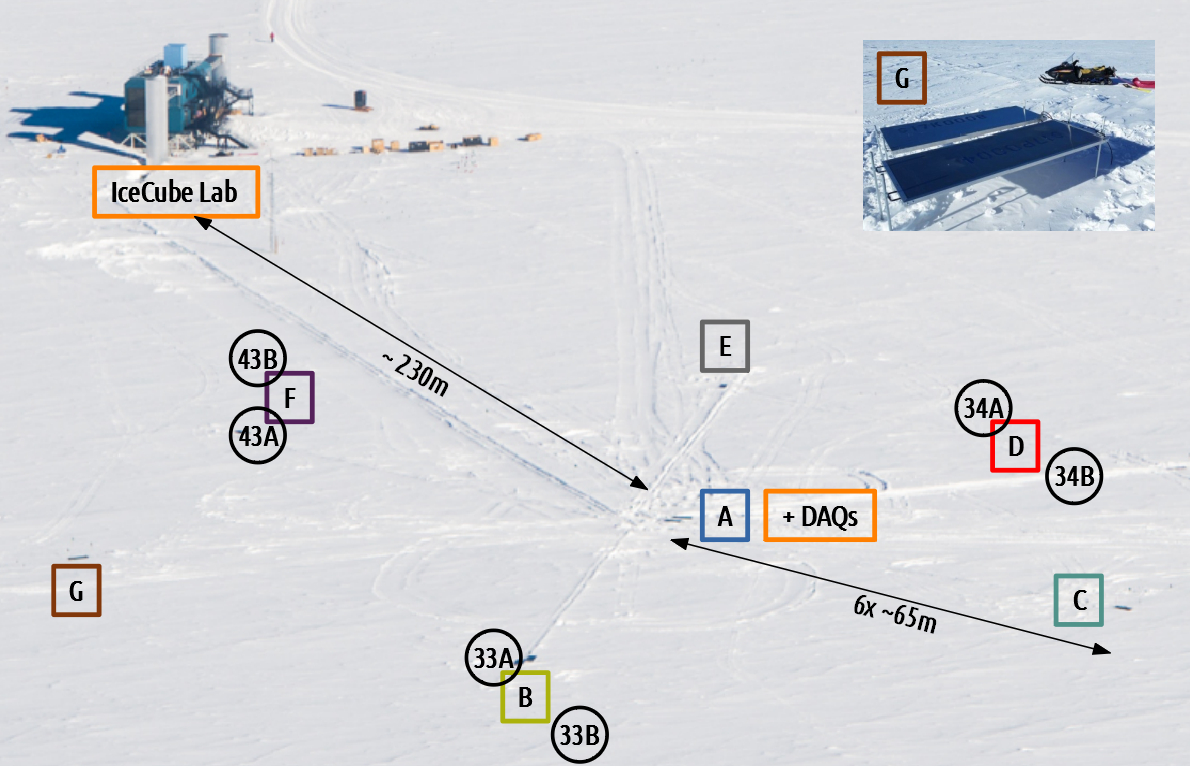}
    \caption{Deployment of the 14 scintillator panels at the South Pole. There are 2 scintillator panels deployed at each location A-G. The nearby IceTop tank positions are shown in circles. The overlapping of the circles with the scintillator locations indicates where the panels are placed on top of an IceTop tank.}
    \label{fig:deployment-pic}
\end{figure}

Accumulating snow cover over the IceTop tanks is continuously increasing the energy threshold for the detection of cosmic ray air showers \cite{Rawlins:2015ztx}. The complex attenuation effects of the snow add systematic uncertainties to air shower measurements, particularly in the mass composition analysis. We have designed and proposed an upgrade to IceTop consisting of homogeneously-spaced scintillator stations with an areal coverage similar to IceTop. We plan on deploying up to 32 scintillator stations over a few years \cite{frank:2019pos, agnieszka:2019pos}. Two prototype stations with seven scintillator panels each were deployed in 2018 shown in Fig.~\ref{fig:deployment-pic}. Each station consists of a different data acquisition (DAQ) architecture and is further discussed in Sec.~\ref{sec:daq}. These two stations provide a proof of design and concept for not only the surface array upgrade but also for the surface field hub concept and White Rabbit timing to be used for IceCube-Gen2.

\section{The Scintillator Panel} 
\label{sec:panels}

Each scintillator detector is designed to have a total sensitive area of 1.5\,m$^2$ and a total weight of less than 50\,kg, to be easily transportable by two people. Each panel comprises 16 extruded plastic scintillator bars (produced by FNAL-NICADD \cite{Beznosko:2005ba}), made of polystyrene with doping of 1\% PPO and 0.03\% POPOP and coated with a 0.25\,$\pm$\,0.13\,mm thick layer of TiO$_2$ reflector. Each bar is 1\,cm thick, 5\,cm wide and 1.875\,m long and has two holes with a diameter of 2.5\,$\pm$\,0.2\,mm. Y-11(300) wavelength shifting fibers (produced by Kuraray\,\footnote{\url{http://kuraraypsf.jp/pdf/all.pdf}}) are routed and looped back through the holes of the bars, resulting in a bundle of 32 fiber ends which is then readout by a 6$\times6$\,mm$^2$ Silicon PhotoMultiplier (SiPM "13360-6025PE", produced by Hamamatsu). More information can be found in \cite{Haungs:2019ylq}.

\section{The Data Acquisition} 
\label{sec:daq}

\begin{wrapfigure}{r}{0.55\textwidth}
    \begin{center}
        \vspace{-30pt}
        \includegraphics[width=0.55\textwidth]{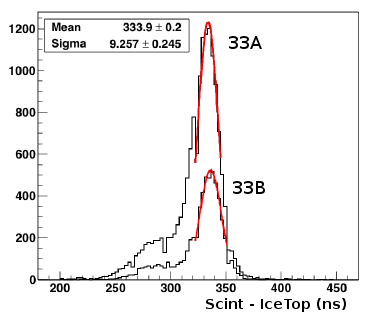}
    \end{center}
    \vspace{-20pt}
    \caption{Timing difference between coincident scintillator and IceTop events.}
    \label{fig:timing}
\end{wrapfigure}

Each of the two prototype stations has a different DAQ concept. One station's DAQ was developed by the Wisconsin IceCube Particle Astrophysics Center (WIPAC) and utilizes digital communications between the scintillator and surface field hub for increased bandwidth. This is known as the microDAQ station. The other station's DAQ was developed by the Deutsches Elektronen-Synchrotron (DESY) and the Karlsruhe Institute of Technology (KIT) and is capable of sending full waveforms back to the surface field hub. This is known as the TAXI station \cite{Karg:2014cra}. Each surface field hub houses White Rabbit GPS timing distribution and communications via fiber link to the IceCube Computing Lab (ICL). Further detailed information about the two DAQs can be found in \cite{Collaboration:2017tdy}.

\section{Scintillator Characterization and Stability} 
\label{sec:stability}
\subsection{White Rabbit Timing} 
\label{sub:wrt}

Verification of GPS timing distribution via White Rabbit was realized using coincident events between IceTop and the scintillators. There are two locations where a scintillator panel is positioned directly above an IceTop tank. The scintillator time delay relative to IceTop of $\sim$335\,ns is measured and shown in Fig.~\ref{fig:timing} and is consistent with 60\,m copper cable delay which is not corrected for online as is with IceTop.

\subsection{Temperature Compensation} 
\label{sub:temp-comp}

\begin{wrapfigure}{r}{0.55\textwidth}
    \begin{center}
        \vspace{-10pt}
        \includegraphics[width=0.55\textwidth]{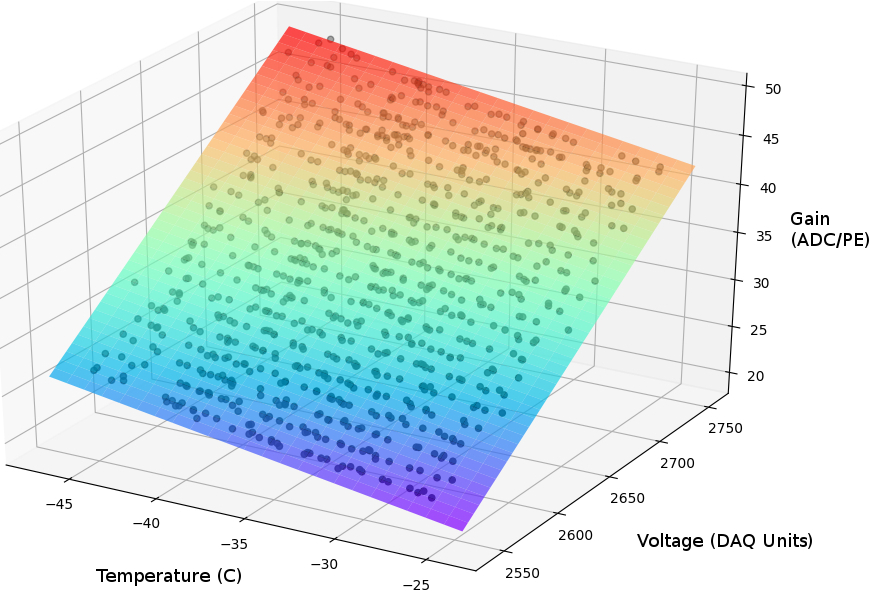}
    \end{center}
    \vspace{-10pt}
    \caption{An example plane fit to the temperature vs voltage vs measured gain of the SiPM.}
    \label{fig:planer-fit}
\end{wrapfigure}

The gain and photo-detection efficiency (PDE) of SiPMs is temperature dependent. The on-board scintillator microDAQ mitigates this effect at the software level by periodically measuring the temperature of the SiPM and adjusting the bias voltage to maintain a constant gain. The temperature dependence of each individual scintillator was mapped out prior to deployment and a plane fit to the temperature-vs-voltage-vs-gain surface is sufficient for the operating range of interest as shown in Fig.~\ref{fig:planer-fit}. The gain is set to 30 ADC/PE (Analog-to-Digital Conversion units per Photo-Electron) to optimize the dynamic range of the microDAQ.

\subsection{Gain and MIP Light Yield}
\label{sub:light-yield}

\begin{figure}[ht]
    \centering
    \includegraphics[width=0.95\textwidth]{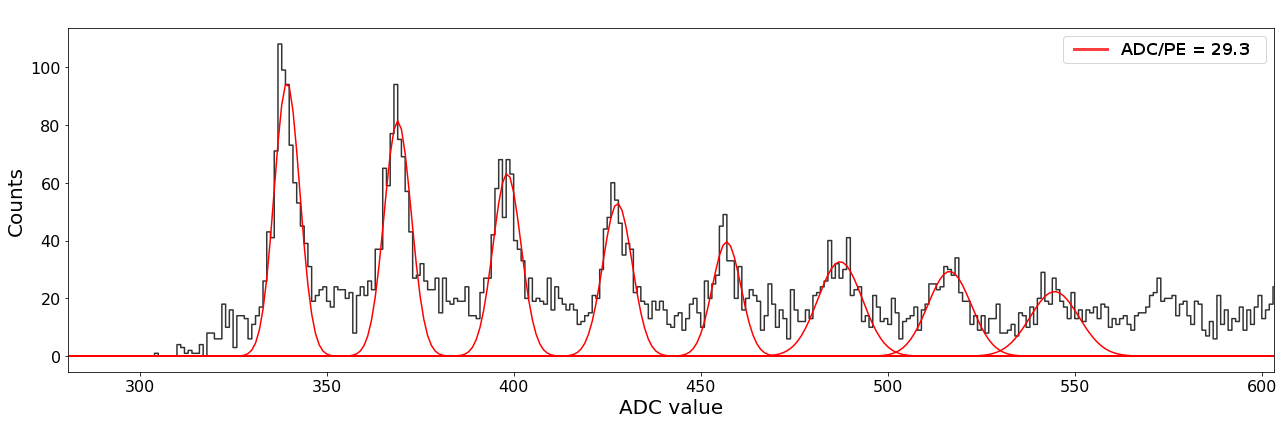}
    \caption{Gaussian fits to the finger spectrum give an average PE separation of 29.3\,$\pm$\,0.4 ADC/PE which is consistent with the FFT result shown in Fig.~\ref{fig:fft-fit}.}
    \label{fig:gauss-fit}
\end{figure}

\begin{figure}[ht]
    \centering
    \includegraphics[width=0.95\textwidth]{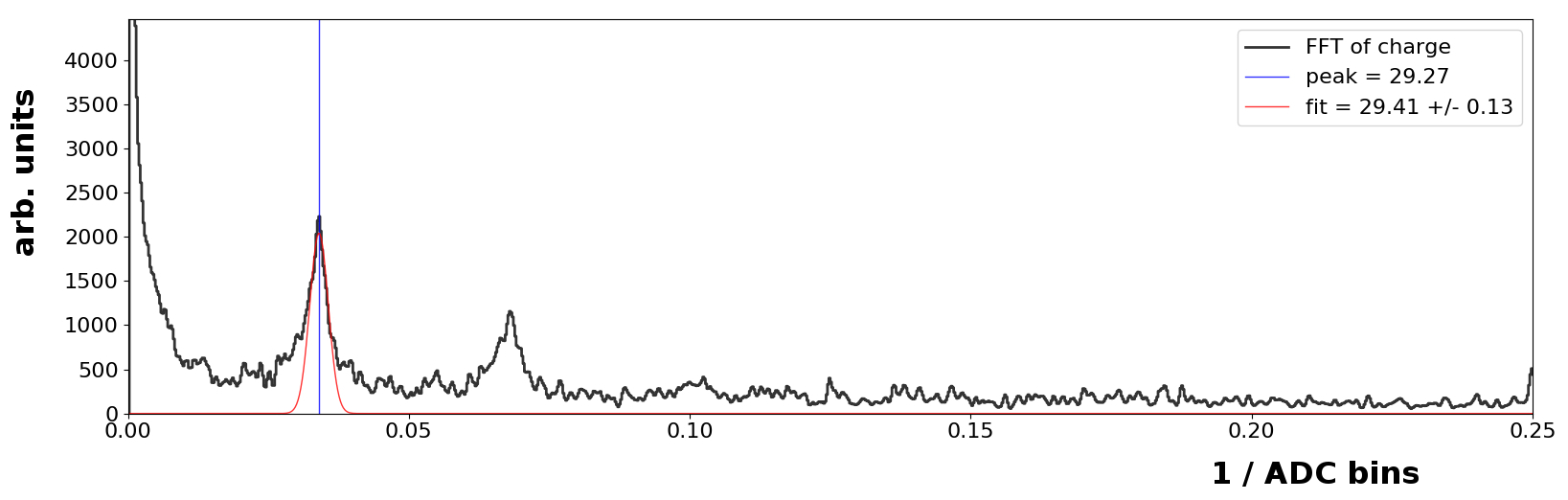}
    \caption{The FFT of charge histogram returns a gain of 29.4\,$\pm$\,0.1 ADC/PE which is consistent with the Gaussian fit method shown in Fig.~\ref{fig:gauss-fit} proving to be a quicker and more robust method for determining the gain of the SiPM.}
    \label{fig:fft-fit}
\end{figure}

The gain of the SiPM is measured and monitored in units of ADC counts vs PE by taking the Fast Fourier Transform (FFT) of the ``finger plot'' charge histogram shown in Fig.~\ref{fig:gauss-fit}. This proves to be a precise and computationally quick method as the finger spectrum is analogous to a sine wave in frequency space as shown in Fig.~\ref{fig:fft-fit}. Comparing the PE separation measured from Gaussian fits in Fig.~\ref{fig:gauss-fit} to the FFT result in Fig.~\ref{fig:fft-fit}, the measurements are consistent. The Gaussian fits give an average gain of 29.3\,$\pm$\,0.4 ADC/PE and the FFT of the charge histogram gives a gain of 29.4\,$\pm$\,0.1 ADC/PE proving to be a quick and robust method for determining the gain of the SiPM in units of ADC/PE. 

\begin{figure}[ht]
    \centering
    \includegraphics[width=0.95\textwidth]{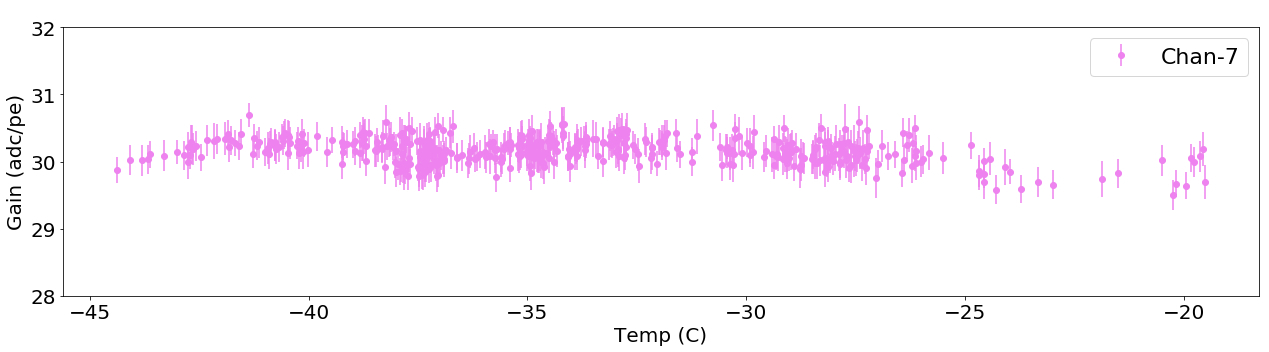}
    \caption{Example temperature-compensated gain stability of one scintillator panel over a broad range of operating temperatures.}
    \label{fig:gain-stability}
\end{figure}

A gain of 30 ADC/PE was chosen to optimize the dynamic range of the microDAQ and 30 ADC/PE is the setpoint for the SiPM gain temperature compensation and is stable within 2\,\% as shown in Fig.~\ref{fig:gain-stability}.

\begin{figure}[ht]
    \centering
    \includegraphics[width=0.95\textwidth]{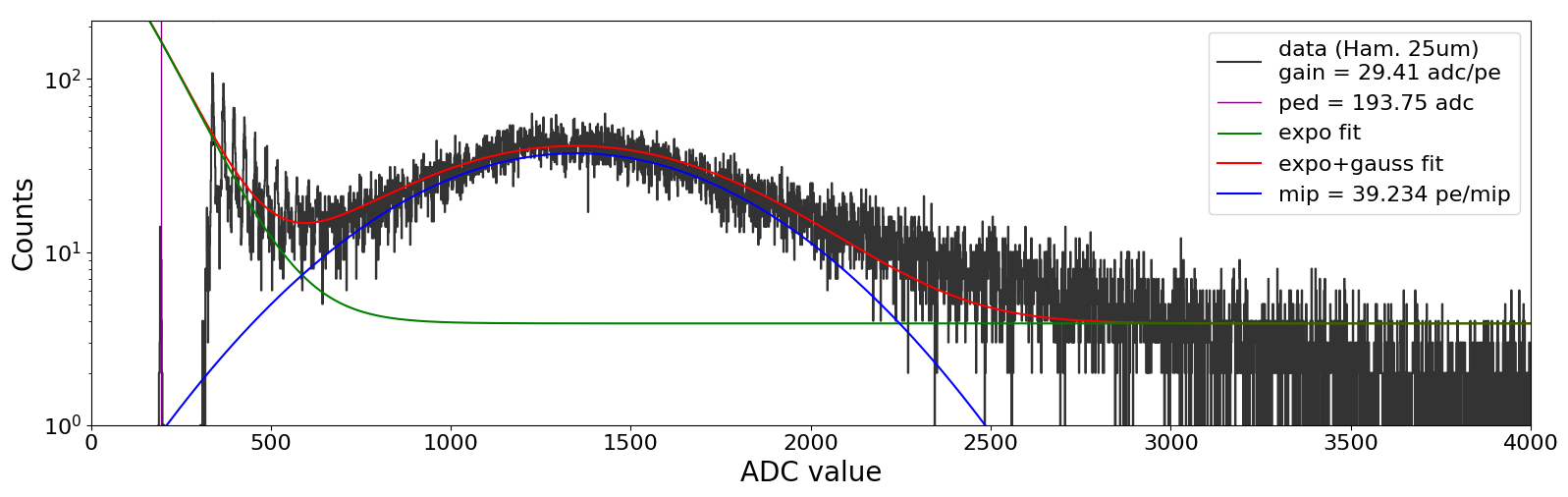}
    \caption{An example fit to the MIP peak. The pedestal is in purple, the exponential dark noise component is green, the MIP component estimated with a Gaussian is in blue, and the fit total is in red. In this example the gain was measured to be 29.4\,ADC/PE and the MIP was fit to 39.2\,PE/MIP. }
    \label{fig:mip-fit}
\end{figure}

The estimated light yield from minimally ionizing particles (MIPs) is determined by an exponential plus Gaussian fit to the spectra after pedestal subtraction and gain correction. An example is shown in Fig.~\ref{fig:mip-fit}. The MIP peak light yield from the scintillators ranges from 39-44 PE/MIP and is consistent with GEANT4 optical simulations \cite{agnieszka:2019pos}.

\subsection{Energy Linearity}
\label{sub:linearity}

\begin{figure}[ht]
    \centering
    \includegraphics[width=0.95\textwidth]{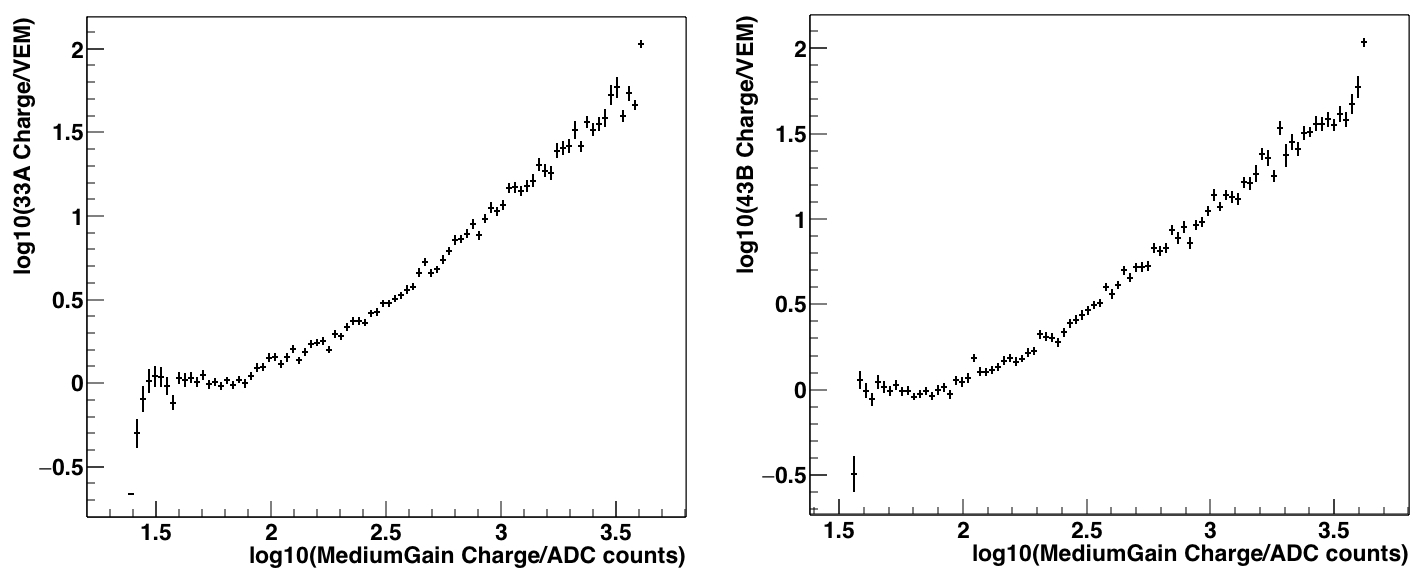}
    \caption{Coincident events are found between IceTop station 33 (tank A) and scintillator location B and IceTop station 43 (tank B) and scintillator location F as shown in Fig.\ref{fig:deployment-pic}. Plots show the charge of scintillator events in ADC counts vs charge of coincident IceTop events in VEM (vertical equivalent muon). }
    \label{fig:icetop-linearity}
\end{figure}

The scintillator charge linearity is evaluated using coincident events with IceTop. We observe a linear charge correlation between IceTop and the scintillators down to IceTop's charge threshold ($\sim$\,0.16\,VEM) as shown in Fig.~\ref{fig:icetop-linearity}. Charge threshold effects in IceTop due to snow coverage are evident and will be significantly improved with the addition of the scintillator surface array \cite{agnieszka:2019pos}.

\section{Performance of the Array} \label{sec:data}

The scintillator data is acquired when one or more scintillator signals cross the discriminator threshold ($\sim$\,~0.2\,MIP). The results shown here require three or more scintillators to have triggered in coincidence ($\pm$\,200\,ns). 

\subsection{Reconstruction of Air Showers} \label{sec:reco}

\begin{figure}[ht]
    \centering
    \includegraphics[width=0.95\textwidth]{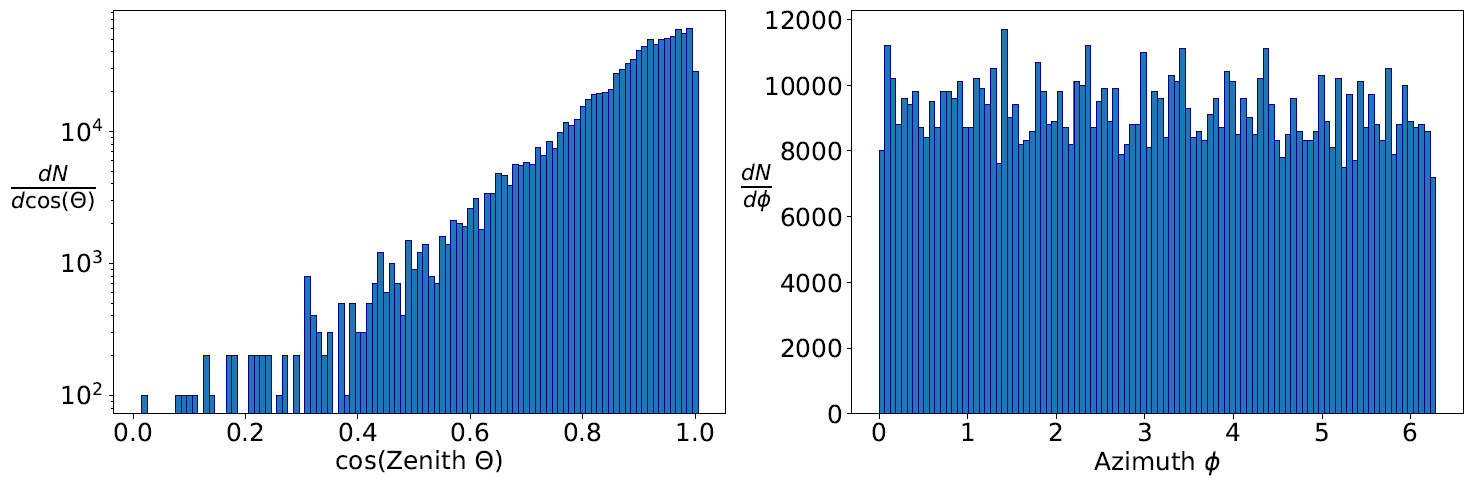}
    \caption{Distributions of the reconstructed zenith and azimuth angles of events detected by the TAXI scintillator station (Sec.~\ref{sec:daq}) which are coincident with IceTop events. }
    \label{fig:scint-reco}
\end{figure}

 The incidence direction of the shower was reconstructed by a plane-front-fit for the scintillator array. The IceCube Laputop algorithm is used for IceTop, whereby the plane-front is calculated and then used as a seed for a fit to the lateral distribution of the signals and shower front curvature. Fig.~\ref{fig:scint-reco} shows the distributions of the reconstructed zenith and azimuth values of the scintillators for events in coincidence with IceTop. 
 
\begin{figure}[H]
    \centering
    \includegraphics[width=0.95\textwidth]{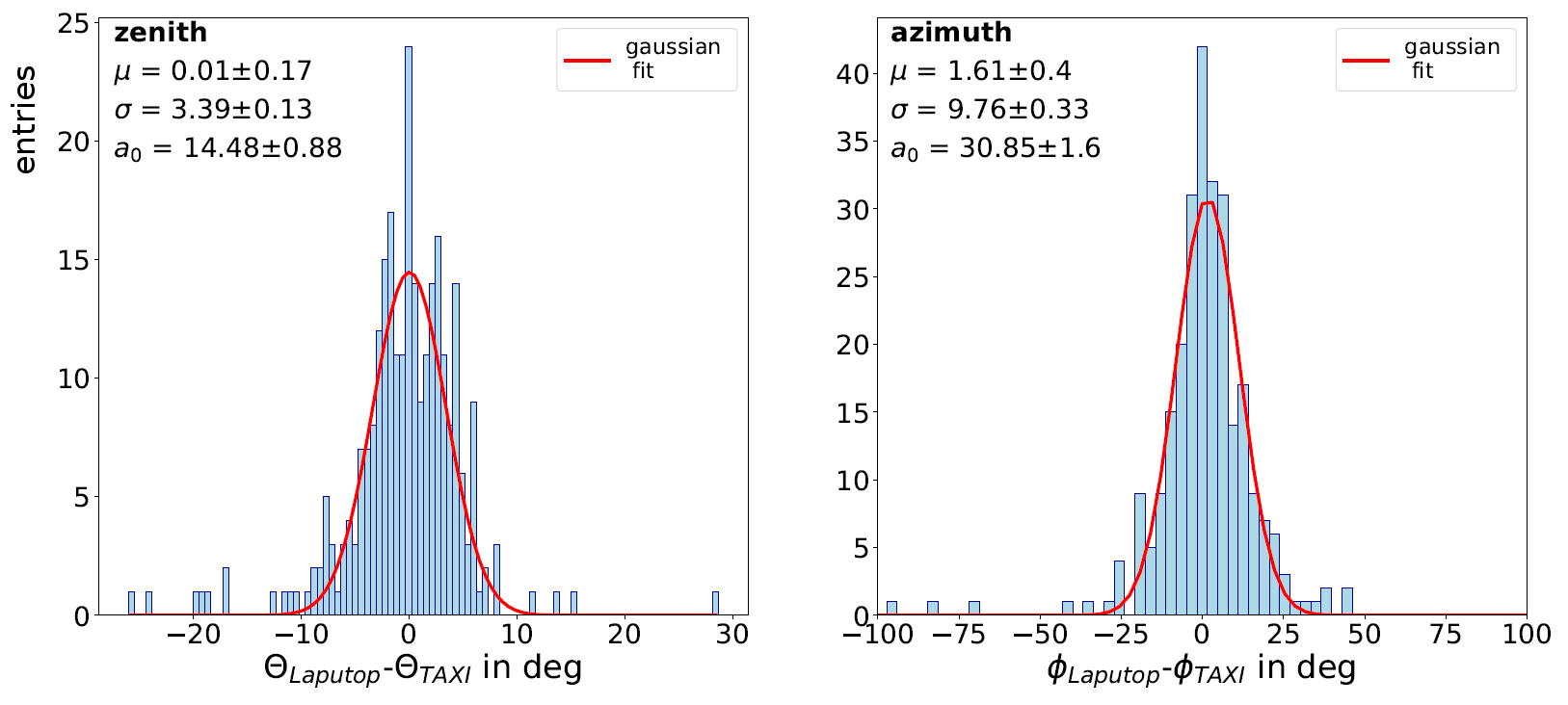}
    \caption{Histograms of the differences between the TAXI scintillator station (Sec.~\ref{sec:daq}) event reconstruction angles and IceTop Laputop reconstructions. Only IceTop events with Laputop reconstructed core positions within the scintillator area are considered. The Gaussian fits are in red with fit results shown in the top-left.}
    \label{fig:scint-icetop-diff}
\end{figure}

The accuracy of the scintillator air-shower reconstruction is estimated by comparing the reconstructed shower plane values with the ones obtained from IceTop data. In Fig.~\ref{fig:scint-icetop-diff} the differences in the zenith and azimuth angle of the plane front between the scintillator array and IceTop Laputop reconstruction are shown. A good correspondence between the results of the reconstructions is obtained, therefore the quality of the timing information is rather good and sufficient for air shower detection.

\section{Summary}

The scintillator upgrade of IceTop is in part a test of the scintillator surface array and in part a test of the infrastructure for the IceCube upgrade. We have demonstrated the stability of the White Rabbit timing distribution and the feasibility of surface field hubs for the upgrade. The scintillator surface array is performing as expected with light yields around 40-45\,PE/MIP as predicted by GEANT4 optical simulations. The charge thresholds of the scintillator surface array are around a factor of five lower than that of IceTop and mitigate the snow accumulation effects observed from IceTop detectors. The direction of incident air showers are reconstructed accurately, taking into account the small number and area of the detectors. 
The deployed scintillator stations are a proof of concept of the feasibility of the detector design and shows it fulfills the requirements of a surface instrumentation upgrade of IceTop.

\bibliographystyle{ICRC}
\bibliography{references}

\end{document}